\documentclass[11pt]{article}

\usepackage[english]{babel}
\usepackage[letterpaper,top=1in,bottom=1in,left=1in,right=1in,marginparwidth=0.75in]{geometry}

\usepackage[version=4]{mhchem} 
\usepackage{siunitx}
\usepackage{graphicx}
\usepackage{dcolumn}
\usepackage{bm}
\usepackage{setspace}
\usepackage{amsmath}
\usepackage{graphicx}
\usepackage{amssymb}
\usepackage{mathtools}
\usepackage{cite}
\usepackage{fixmath}

\title{Digital autofocusing of a coded-aperture \\ Laue diffraction microscope}
\author{Do\u ga G\" ursoy\footnote{E-mail: dgursoy@anl.gov}, Dina Sheyfer, Michael Wojcik, Wenjun Liu, Jonathan Z. Tischler \\
X-ray Science Division, Argonne National Laboratory, \\
9700 S Cass Ave, Lemont, IL 60439, USA}

\begin{document}
\maketitle
\doublespacing

\begin{abstract}
To provide optimal depth resolution with a coded-aperture Laue diffraction microscope, an accurate position of the coded-aperture and its scanning geometry need to be known. However, finding the geometry by trial and error is a time-consuming and often challenging process because of the large number of parameters involved. In this paper, we propose an optimization approach to automate the focusing process after data is collected. We demonstrate the robustness and efficiency of the proposed approach with experimental data taken at a synchrotron facility. 
\end{abstract}

\section{Introduction}

Laue diffraction imaging by scanning a sample with a sub-micrometer focused beam from a synchrotron radiation source has been an established technique to study the microstructural defects, strain and deformations in crystalline materials and their relation to the physical properties at macroscopic scales \cite{ice20003d, kunz2009dedicated, tamura2002submicron, ulrich2011new, hofmann2009probing, maabeta2006defect, park2007local}. In a typical Laue diffraction microscope, we scan a crystalline sample point by point across a grid of positions while recording the diffracted x-rays by a pixel-array area detector for each sample position. Because individual crystallites along the illumination path diffract the incident beam of polychromatic x-rays passing through the sample, analysis of the recorded data reveals the diffraction information about the local structure of the crystal lattice and associated strain and orientations. 

\begin{figure}
\centering
\includegraphics{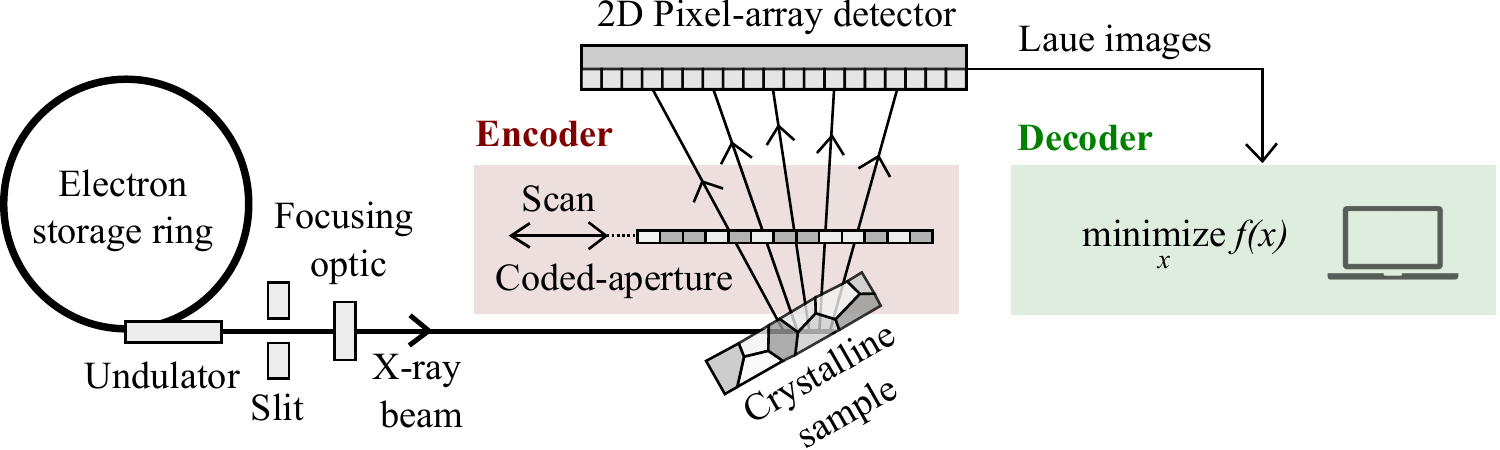}
\caption{The schematic representation of a coded-aperture Laue diffraction microscope. A crystalline sample is illuminated by focused x-ray beam produced by synchrotron radiation available from wiggler or undulator sources. Laue diffraction from crystallites is recorded using a pixel-array detector mounted in a \SI{90}{\degree} reflection geometry. Scanning the coded-aperture enables an effective encoding of the signals diffracted from different depths. The recorded signals are decoded digitally by an optimization algorithm to resolve the Laue patterns and their originating depths along the beam direction.}
\label{fig1}
\end{figure}

In a Laue diffraction microscope, the depth information along the illumination path for a given beam position can be obtained by passing an absorbing structure across the diffracted Laue beams. The most straightforward and common approach is to scan a highly absorbing wire or a sharp structure as a differential-aperture to resolve the depth of the beams one by one \cite{Larson:02}. The method works similar to a scanning pinhole-camera in which only a beam of rays originating from a point will be allowed into the camera and the origin of all the beams can be obtained by scanning. As a time-efficient alternative, we can use coded-apertures, a well established technique in x-ray astronomy \cite{dicke1968scatter}, to accelerate the data acquisition process. Coded-apertures has a long history with applications ranging from spectroscopy \cite{golay1949multi} to imaging \cite{mertz1965transformations}, and we have recently demonstrated use of coded-apertures with pink beam as a potential imaging tool for studying crystalline materials \cite{Gursoy2022coded}. In a coded-aperture Laue microscope, we scan the coded-aperture while collecting Laue images of a pink beam diffracted from a polycrystalline sample (Fig.~\ref{fig1}). Scanning a patterned structure such as a coded-aperture allows controlled modulation of recorded intensities, from which we can resolve Laue diffraction as a function of depth along the beam by using computational methods.

Achieving the resolution limit of a coded-aperture Laue microscope requires both an accurate characterization of the coded-aperture and knowledge of its geometry during scanning. Fabricating a millimeter sized coded-apertures with micrometer features is challenging and usually the structure of the fabricated coded-aperture deviates from the assumed structure. We can use optical or scanning electron microscopes to take images of the aperture surface and its sides, but acquiring its shape in 3D such as its thickness or in-plane bending radius is difficult. In addition, because the field-of-view of the images is much smaller than the size of the coded-aperture, we need to take multiple images and stitch them afterwards to obtain the final structure. This process can induce artifacts due to such as shallow depth-of-field of the microscopes when the coded-aperture is not perfectly planar. Similar to the characterization of the 3D structure of the coded-aperture, the positioning and scanning geometry of the coded-aperture before the experiment is not always available unless a tailored metrology system, e.g. with laser interferometers that can track the positioning of the coded-aperture in 3D with respect to the sample and the beam, is developed. When a metrology system is not available, we use a highly fluorescent sample such as a copper plate and record the emitted x-ray fluorescence signals to have a rough estimate about the position of the coded-aperture. However, the accuracy of this calibration process is relatively low (the misalignment error can reach up to \SI{5}{\micro\meter}) because of the noisy measurements of the fluorescence emission.

A more viable approach is to estimate positioning parameters and its geometry digitally after the data is collected. To this end, we propose a digital autofocusing method that can provide the geometry with a high accuracy. Our approach uses data from a known calibration sample such as a defect-free thin crystal, and numerical optimization to adjust the focusing parameters of the coded-aperture until all the resolved signals originate from their assumed depths.

\section{Methods}

In this section, we describe the geometrical model for the coded-aperture Laue diffraction microscope, an algorithm for digital autofocusing of the depth-resolved signals, and our experimental setup. 

\subsection{Basics of coded-aperture Laue diffraction imaging}

We first introduce the basics of coded-aperture imaging and the mathematical basis before presenting our digital autofocusing approach. Assume that the coded-aperture is represented by the vector $\mathbold{a} = [a_1, a_2, \hdots, a_L]^T$ with each coefficient describing the optical transmissivity of the coded-aperture along incident direction of the beam, and $L$ is the number of components of its piece-wise constant representation. For example, a $0$ means that an incident beam on the coded-aperture can pass through without being absorbed and a $1$ means that the signal is fully absorbed. In a realistic setting, the values are often between $0$ and $1$, and are dependent on the thickness and material of the coded-aperture, as well as the energy and incidence angle of the incident beam. The interaction of the beam with a segment of the coded-aperture $\mathbold{a}_p = [a_{p}, \hdots, a_{p+N-1}]^T$ can be expressed with a matrix product as follows,
\begin{equation}
\label{eq1}
\begin{bmatrix}
    d_{1} \\
    d_{2} \\
    d_{3} \\
    \vdots \\
    d_{M}
\end{bmatrix}
=
\begin{bmatrix}
    a_{p} & a_{p+1} & a_{p+2} & \dots  & a_{p+N-1} \\
    a_{p+1} & a_{p+2} & a_{p+3} & \dots  & a_{p+N} \\
    a_{p+2} & a_{p+3} & a_{p+4} & \dots  & a_{p+N+1} \\
    \vdots & \vdots & \ddots & \ddots & \vdots \\
    a_{p+M-1} & a_{p+M} & a_{p+M+1} & \dots  & a_{p+M+N-2}
\end{bmatrix}
\begin{bmatrix}
    s_{1} \\
    s_{2} \\
    s_{3} \\
    \vdots \\
    s_{N}
\end{bmatrix},
\end{equation}
where $\mathbold{d} = [d_1, d_2, \hdots, d_M]^T$ is the intensity vector to a single detector pixel that we acquire by translating the coded-mask $M$ times, and $\mathbold{s} = [s_1, s_2, \hdots, s_N]^T$ is the signal along the illumination path scattered into this pixel. We can use matrix notation to express Eq.~\ref{eq1} as,
\begin{equation}
\label{eq2}
    \mathbold{d} = \mathbold{A}_p \mathbold{s},
\end{equation}
where $\mathbold{A}_p$ is the coding matrix of size $M \times N$. A feasible $p$ and $\mathbold{s}$ satisfying Eq.~\ref{eq2} can be obtained by solving the following minimization problem,
\begin{equation}
\label{eq3}
    \min_{p, \mathbold{s}} \left\|  \mathbold{A}_p \mathbold{s} - \mathbold{d} \right\|_2^2,
\end{equation}
which requires jointly resolving the signals and their intersecting positions with the coded-aperture during scanning. We can use a sequential optimization approach in which we update $p$ by fixing $\mathbold{s}$, followed by an update of $\mathbold{s}$ by fixing $p$. Once all of the signals and their positions are recovered, we can use ray-tracing to register these to their corresponding depths along the illumination path.

One key remark is that the energy of the diffracted beam is unknown, so we only know the relative changes in coefficients, but not their absolute values. Therefore, we use normalized coding matrix by constraining the dynamic range of the coefficients between 0 and 1. As a consequence, we normalize data such that $\bm{d} = (\bm{d}_{raw}-\mu_0)/(\mu_1-\mu_0)$ using the average intensities corresponding to the 1s and 0s in the raw scan data $\bm{d}_{raw}$. Because we don't have direct access to the average intensities, we estimate them from the minimum and maximum intensities in the scan data such that $\mu_0 = d_{min} + 2\sqrt{d_{min}}$ and $\mu_1 = d_{max} - 2\sqrt{d_{max}}$, where $2\sqrt{d_{min}}$ or $2\sqrt{d_{max}}$ correspond to two standard deviations from the mean. When the incident intensity is unstable, we can consider using an additional measurement setup to monitor the incident beam and weight the measurements proportionally as part of the normalization procedure.

\subsection{Geometrical model and its parameterization}

\begin{figure}
\centering
\includegraphics{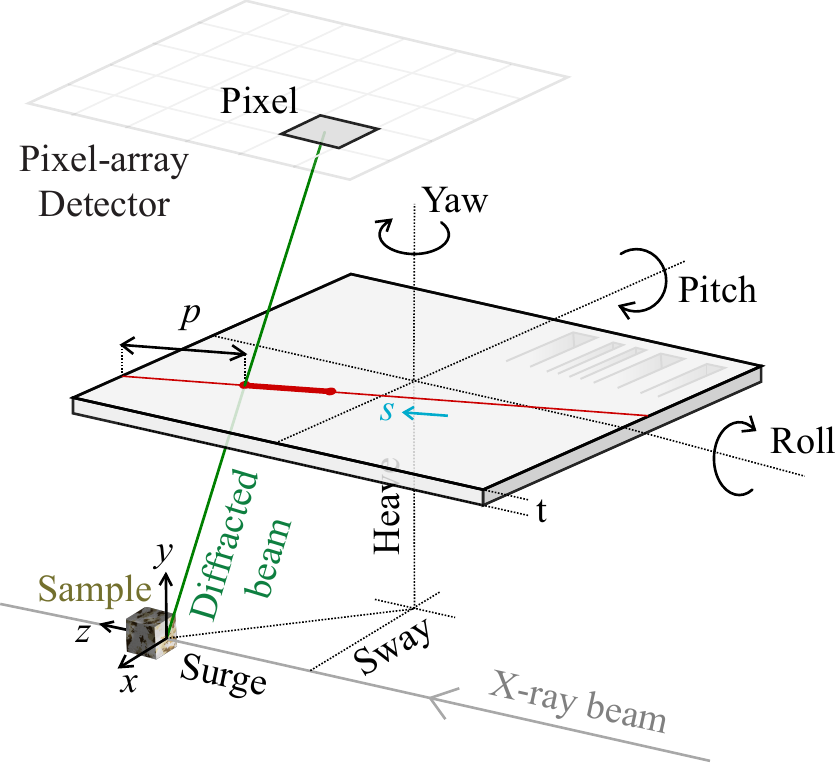}
\caption{The sketch of the geometrical model for positioning and scanning of the coded-aperture. The unit vectors  $x$, $y$ and $z$ define the laboratory coordinates in which the position of the detector is calibrated with respect to the sample. For positioning the coded-aperture, we use six degrees of freedom (6DOF) that describes the movements up/down along the $y$ axis (heave), forwards/backwards along they $z$ axis (surge), left/right along the $x$ axis (sway), and the changes in orientation through rotation about normal (yaw), transverse (pitch), and longitudinal (roll) axes. The scalar $p$ shows the initial intersection position of the diffracted beam with the coded-aperture along the intersection path (red line).}
\label{fig2}
\end{figure}

We describe the position of the coded-aperture in the laboratory coordinate system in which the detector position is calibrated. The detector is mounted in a \SI{90}{\degree} reflection geometry and its position is calibrated with respect to a fixed thin sample (e.g., \SI{10}{\micro\meter} \ce{Si}) located at the origin of the laboratory coordinates (see, Chapter 2 and 10 in \cite{barabash2014strain} of for details in detector calibration). For example, \ce{Si} with thickness in the range of 1-\SI{10}{\micro\meter} is quite stable and can stay stationary while collecting data. A graphical representation of the geometrical model is given in Fig.~\ref{fig2}. The coded-aperture has a thickness of $t$ and is free to change position as forward/backward (surge), up/down (heave), left/right (sway), as well as its orientation about three perpendicular axes intrinsic to the coded-aperture (yaw, pitch and roll). We also illustrate in Fig.~\ref{fig2} a signal diffracted into an arbitrary detector pixel and its intersection path with the coded-aperture along the scan direction. The path starts at position $p$ and its apparent length is variable for each pixel dependent on the geometry. Because the patterning of the coded-aperture along the intersection path performs the encoding, we need to know the geometry to accurately construct the coding matrix from this pattern and solve the problem in Eq.~\ref{eq3}. 

\subsection{Effect of geometry on the encoding}

\begin{figure}
\centering
\includegraphics{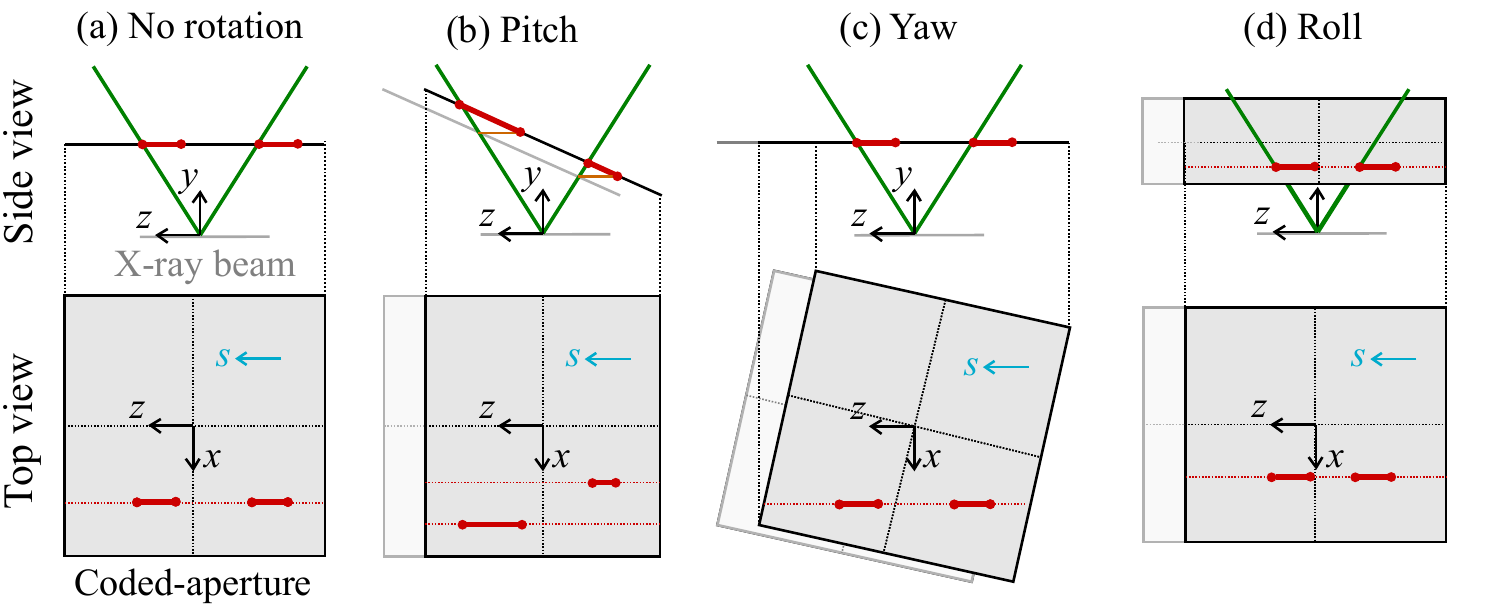}
\caption{Illustration of the effect of the coded-aperture rotations on encoding of measurement signals (a) without rotation and with nonzero (b) pitch, (c) yaw, and (d) roll angles. Both the side and top views of the microscope is shown. The segments of the coded-aperture that performs the encoding (shown in red) can vary depending on the geometry. }
\label{fig3}
\end{figure}

Before we introduce our digital autofocusing algorithm, we would like to make some remarks on the effect of coded-aperture geometry on the encoding in recorded intensities. In an ideal setting, we aim to acquire data such that the coded-aperture lies in parallel to the xz-plane with its bars perpendicular to the scanning axis. In this geometry, scanning the coded-aperture along a scan direction in xz-plane provides an identical encoding for all pixel measurements and constructing the coding matrix in Eq.~\ref{eq3} is straightforward. However, in reality, it is challenging to manually align the coded-aperture with respect to the xz-plane at a sub-micrometer precision, and it may experience small deviations from the ideal geometry. To understand the effects of geometry on the encoding, we study in Fig.~\ref{fig3} the positions of a coded-aperture before and after its translation along its scan direction for three basic rotations. The segment of the coded-aperture that modulates the diffracted beam can change depending on the rotation type, which leads to a slightly distorted encoding relative to the assumed geometry. Especially, a nonzero pitch and yaw of the coded-aperture lead to widening or shortening of the segments, and therefore they affect the encoding. Because the bars of the coded-aperture are aligned with the x-axis, a rolling has zero effect on the encoding unless it is combined with a pitch or yaw motion. While the conceptual drawings in Fig.~\ref{fig3} provide valuable insights about anticipated changes in the modulation, we quantify the errors in the results section as part of the sensitivity analysis.

\begin{figure}
\centering
\includegraphics{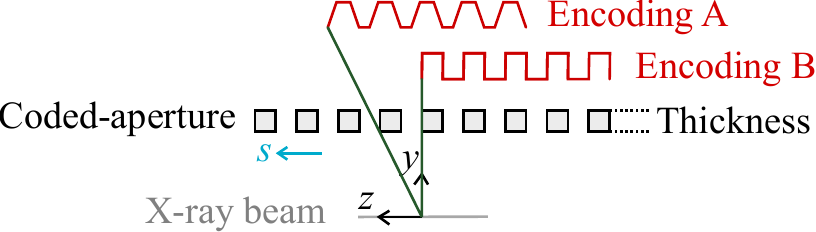}
\caption{Illustration of the effect of the coded-aperture thickness on encoding. Two diffracted beams (shown in green) with different angles of incidence on the coded-aperture experience a different absorption profile and they yield different encodings.}
\label{fig4}
\end{figure}

The thickness of the coded-aperture along its optical axis also contributes to the encoding in measurements, see Fig.~\ref{fig4}. Because the absorption length for the diffracted beam is changing with the incidence angle of the beam on the coded-aperture, recorded intensities become more convoluted with increasing incidence angle and with increasing thickness. For example, the standard setup at the 34-ID-E beamline of Advanced Photon Source (APS) uses a $409.6\times\SI{409.6}{\milli\meter}^2$ pixel-array detector mounted vertically \SI{51.2}{\centi\meter} above the sample. This setup is expected to experience diffraction up to an incidence angle of about \SI{20}{\degree} and can considerably distort the reconstruction of signals, especially when the aspect ratio of the coded-aperture is high. In addition, when we use multiple detectors to cover a larger solid angle, the incident angle can get larger. Therefore, modeling the thickness in addition to its position and orientation is critical in constructing the coding matrix for an accurate estimation of its geometry and depth-resolved signals.

\subsection{Calibration data requirements for autofocusing}

Our digital autofocusing approach requires a calibration dataset recorded in a controlled setting for estimation of the position (surge, sway, heave) and the orientation (yaw, pitch, roll) of the coded-aperture. As the calibration data to perform digital autofocusing, we use the same setup to calibrate the detector and collect a series of Laue images from a known sample such as a \SI{10}{\micro\meter} strain-free \ce{Si} crystal while translating the coded-aperture along the z-axis. We select a sufficiently large scan range to obtain a dataset with sufficient redundancy. The range or the exposure time can be shortened or widened based on the source brightness aiming to have signals above a certain signal-to-noise ratio (SNR). We choose the step size for the coded-aperture close to the desired depth resolution. For example, when we aim to resolve the signals from a \SI{100}{\micro\meter} thick sample at \SI{1}{\micro\meter} resolution, we need to scan our coded-aperture with a \SI{1}{\micro\meter} step size. To target an isotropic spatial resolution across the volume, a step size matching the diameter of the focusing beam can be used. 

\subsection{Digital autofocusing approach}

\begin{figure}
\centering
\includegraphics{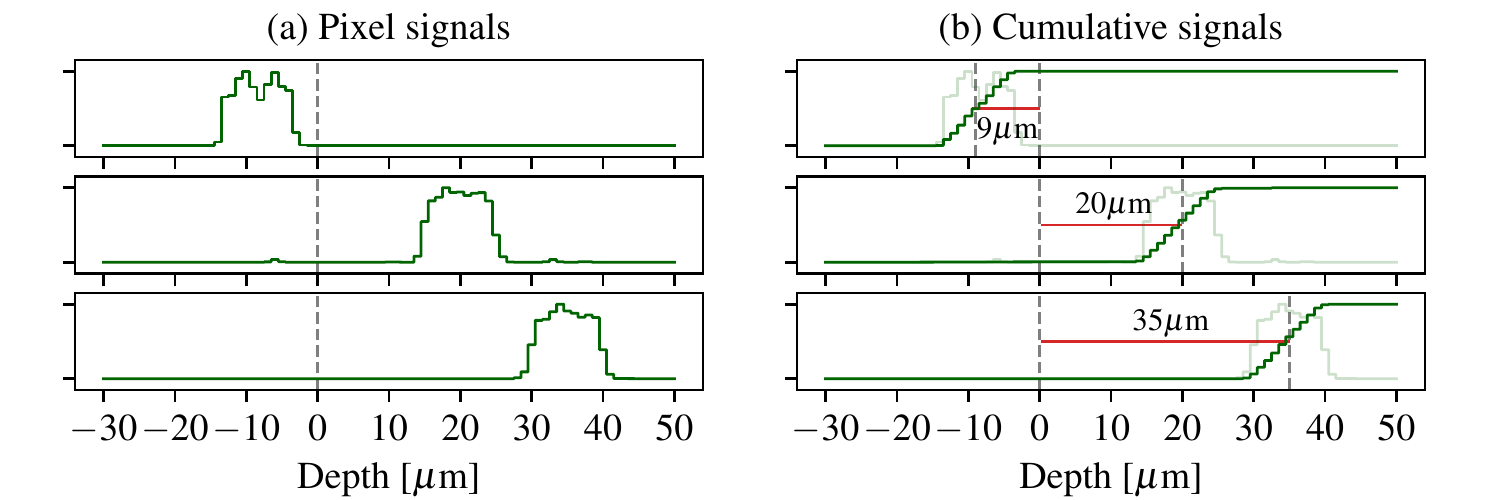}
\caption{(a) Depth-resolved signals for three selected pixels for an arbitrary coded-aperture geometry. Because the geometry is incorrect, these signals are out of focus. (b) Cumulative sum plots associated with the plotted depth-resolved signals. The positions corresponding to the median values of the signals and their distances from the origin are marked with dashed lines. The digital autofocusing uses an optimization routine to minimize these distances, or equivalently, the variance of the resolved signal positions.}
\label{fig5}
\end{figure}

From the collected calibration dataset, our goal is to find parameters such that the depth-resolved signals recorded in each detector pixel are all focused on the assumed position of calibration sample. One way to achieve this is by minimizing the variance of the recovered positions of the depth-resolved signals with respect to the origin of the laboratory coordinate system,
\begin{equation}
    \min_{\mathbold{g}} \: \operatorname{Var}(\mathbold{p}_\mathbold{g}),
    \label{eq4}
\end{equation}
where $\mathbold{g}$ is the unknown geometry vector (e.g., $[g_\text{surge}, g_\text{sway}, g_\text{heave}, g_\text{yaw}, g_\text{pitch}, g_\text{roll}]^T$) that we want to optimize, $\mathbold{p}$ is the vector holding the recovered signal positions with respect to the origin, and $\operatorname{Var}(\mathbold{p})$ is its variance. In Fig.~\ref{fig5}, we explain how we compute the position vector $\mathbold{p}$. We resolve every signal corresponding to the selected pixels and use their positions in which the median intensity is located to compute the position vector. For the solution of problem in Eq.~\ref{eq4}, we use the coordinate-descent method \cite{ortega2000iterative} that successively minimizes along coordinate directions of the geometry vector. 

Note that the number of pixels used in the analysis sets the size of the position vector. For a reliable variance estimate, we can pick the pixels with high signal-to-noise ratio (SNR) and neglect the remaining pixels in the optimization. A good practice is to use the highest allowable exposure without saturating the detector and use the pixels around the Bragg peaks. For example, with a 16-bit dynamic range detector, we can achieve a maximum SNR of $2^{16}/\sqrt{2^{16}} = 256$. In this setting, usually the pixels in the neighborhoods of a Bragg peak also yield decent SNR. In our experiments, the SNR of the pixels range from 12 to 195. Besides, using a fraction of pixels for the analysis is favorable to reduce the computational need for solving Eq.~\ref{eq4}. However, we can also increase the size of the position vector by collecting more data by increasing the scan range. With this approach, we can leverage the redundancy in dataset by separating the scan data into different bins each having a different range of scan points. This increases the size of the position vector, and in turn, can improve the optimization process by including more signal positions in the variance estimation. 

\subsection{Data acquisition setup}

\begin{figure}
\centering
\includegraphics{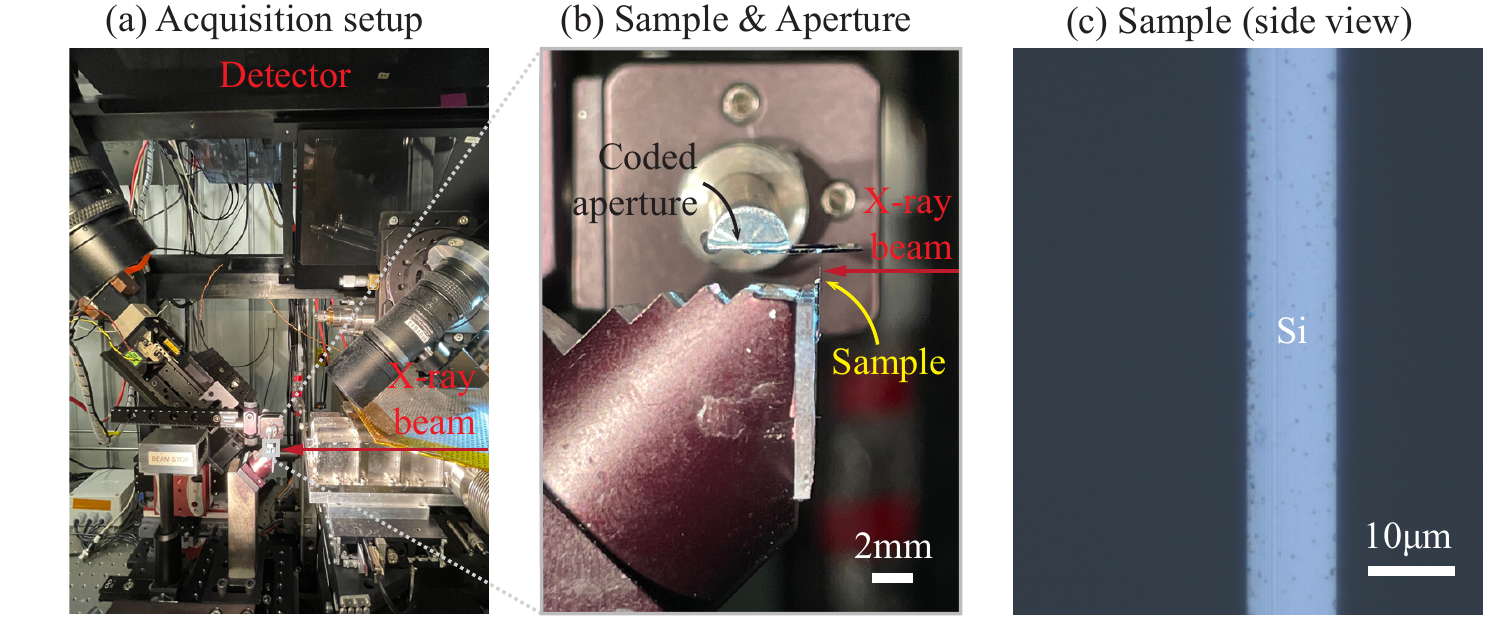}
\caption{(a-b) Photographs of the data acquisition setup at the 34-ID-E beamline of APS at Argonne National Laboratory. (a) Zoomed-out view of the setup that shows the detector on top, and (b) zoomed-in view of the sample and the coded-aperture. The coded-aperture is scanned along the x-ray beam direction to modulate the diffraction coming from the \SI{10}{\micro\meter} thick \ce{Si} sample that is placed perpendicular to beam direction. (c) The image of the sample taken with an optical microscope at $\times$100 magnification.}
\label{fig6}
\end{figure}

We collected our calibration dataset and performed experiments at the dedicated Laue diffraction microscopy setup at the 34-ID-E beamline of APS at Argonne National Laboratory (Fig~\ref{fig6}). The setup used a pair of non-dispersive Kirkpatrick-Baez mirrors to focus a polychromatic x-ray beam with energies in the range of 7-\SI{30}{\kilo\electronvolt} to a spot of about \SI{300}{\nano\meter} diameter. We used a pixel-array Perkin-Elmer detector ($409.6\times\SI{409.6}{\milli\meter}^2$, $2048\times2048$ pixels, and 16-bit dynamic range) mounted in a \SI{90}{\degree} reflection geometry \SI{510.9}{\milli\meter} above the sample for acquiring the Laue patterns. We designed our coded-aperture from a de Bruijn sequence \cite{DeBrujin:46} of length 256 and order 8. De Bruijn sequences refer to a class of nonlinear cyclic sequences, in which every possible string of a particular length occurs exactly once as a substring. We use a greedy algorithm \cite{gabric2018framework} for generating the coding sequence. We manufactured our coded-aperture in the cleanroom at the Center for Nanoscale Materials at Argonne National Laboratory using direct-write lithography. It had \SI{7.5}{\micro\meter} thickness of gold (\ce{Au}) for the 1-bits and a \SI{15}{\micro\meter} thickness of silicon nitride (\ce{Si3Ni4}) membrane with negligible absorption for the 0-bits. We measured the thickness of the coded-aperture along its optical axis as $\SI{4.6}{\micro\meter}$ using a scanning electron microscope.

\section{Results}

In this section, we present a sensitivity analysis for drawing an insight about parameters, results from the digital autofocusing with calibration data, and validate the results with experimental data collected from a highly deformed \SI{75}{\micro\meter} \ce{Ni} sample.

\subsection{Sensitivity analysis}

To understand the effect of changes along coordinate directions on the solution, we first study the cost functions at different geometries of the coded-aperture as depicted previously in Fig.~\ref{fig2}. These observations will provide a basis to set the algorithm parameters such as initial conditions or bounds to the solution search space, and to design a tailored algorithm for the technique. Because the sharpest signals can be obtained when the incident beam is perpendicular to the sample surface, as the calibration sample, we use a defect-free single-crystal silicon (\ce{Si}) membrane with a thickness of about \SI{10}{\micro\meter}, which we mount at a \SI{90}{\degree} angle towards the incident x-ray illumination. The x-ray spot that we used was \SI{100}{\micro\meter} below the edge of the sample. We place the coded-aperture about \SI{1}{\milli\meter} above the sample and aligned it based on the x-ray fluorescence signal from a copper plate such that the bars are roughly perpendicular to the scanning direction. In this geometry, the coded-aperture can simultaneously modulate all signals originating from the sample to the detector pixels through scanning. Also when the sample doesn't block the path of the coded-aperture, we can collect more data by scanning a longer length. To this end, we collected 2000 scan data points by scanning it with a \SI{1}{\micro\meter} step size along the beam direction over a distance of \SI{2}{\milli\meter}. 

\begin{figure}
\centering
\includegraphics{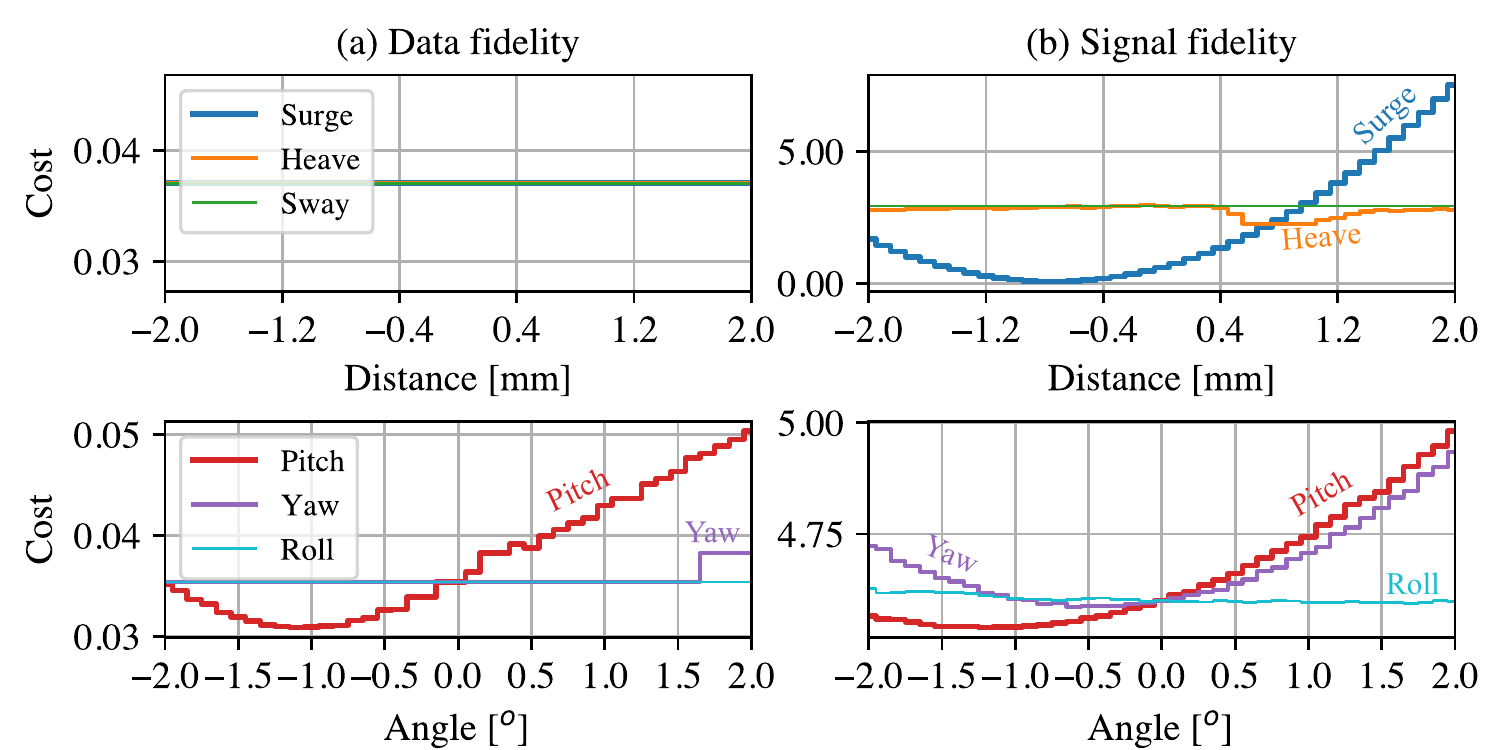}
\caption{The cost functions of the position and orientation of the coded-aperture with respect to different coordinate axes. (a) The data fidelity term in Eq.~\ref{eq3}. (b) The signal fidelity term in Eq.~\ref{eq4}}.
\label{fig7}
\end{figure}

Fig.~\ref{fig6} shows the data fidelity and signal fidelity terms corresponding to the cost functions in Eq.~\ref{eq3} and Eq.~\ref{eq4}. As we expect, the position (surge, heave, sway) of the coded-aperture has no effect on the data fidelity term, because it has no effect on the length of the coded-aperture segment that modulates the diffracted beams. The data fidelity is most sensitive to pitching of the coded-aperture. A nonzero yaw angle also affects the data fidelity, but its contribution is insignificant for angles in the range of a few degrees. Also note that the cost functions in Eq.~\ref{eq4} is dependent on the solution of Eq.~\ref{eq3} and vice versa, because the geometry of the coded-aperture affects the recovery of signals and the recovered signals affect the geometry optimization. Therefore, optimizing the pitch angle before optimizing along other coordinates can yield a faster convergence. 

When we perform ray-tracing to the reconstructed signals, we can compute the cost function in Eq.~\ref{eq4}. The signal fidelity is the most sensitive to the position of the coded-aperture along z-axis (surge), because the surge length determines the position of the focus along z-axis. Optimizing the surge length places the focus roughly at the origin of the laboratory coordinate system and can be optimized prior to optimizing position along x and y axes. 

Another observation we make is the smooth and quadratic nature of the cost functions. Therefore, we prefer using a binary search algorithm, which is a relatively fast search algorithm with a logarithmic runtime complexity  over the search interval for each coordinate-descent iteration.  We perform a sensitivity analysis as in Fig.~\ref{fig7} to specify a set of boundary values that define a feasible search interval for each coordinate. Usually a search space of a few degrees and a few hundreds of micrometers is sufficient when the coded-aperture is roughly aligned with an optical microscope. The binary search algorithm starts comparing the cost functions at the boundary points with the one at the middle point of the interval. If the cost function value at the middle is closer to one of the cost functions at the boundary, the half in which the minimum value cannot lie is eliminated and the search continues on the remaining half of the search interval, again by computing the cost function at the middle point to compare with the boundary pints of the current interval. The process repeats until a desired error tolerance is reached.

\subsection{Digital autofocusing with calibration data}

\begin{figure}
\centering
\includegraphics{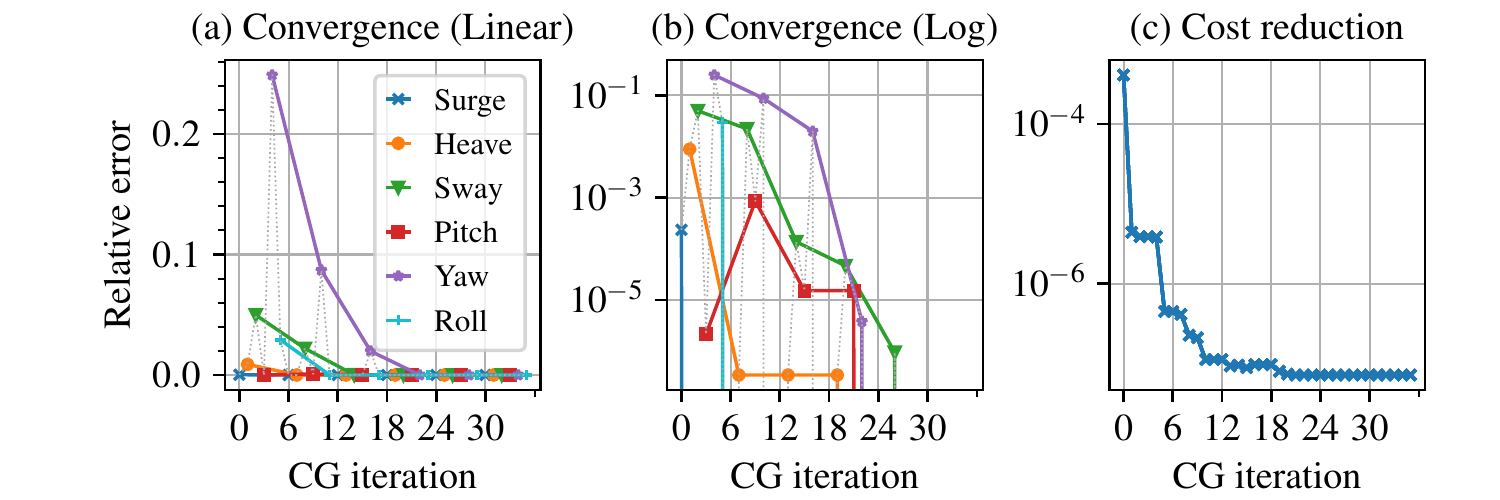}
\caption{Convergence plots of the coordinate-descent (CG) algorithm showing the relative error in linear (a) and log (b) scales in geometrical parameters between successive CG iterations, and (c) the cost values for each CG iteration.}
\label{fig8}
\end{figure}

For testing the algorithm we separated the 2000 scan points into 5 bins each using a scan range of \SI{400}{\micro\meter}. Because of the intensity fluctuation in some of the binned data, we only use the first (0--\SI{400}{\micro\meter}) for calibration and the last bin (1600--\SI{2000}{\micro\meter}) for validation, but using different ranges and different number of bins will probably yield similar results. The selection of the bins and the range depends on the data acquisition conditions. For example, we could chose a larger bin size if the beam is stable during the course of the scan for this bin. If the beam is unstable, we can monitor the intensity and possible drifts during acquisition and cherry pick bins that are free of artifacts or unexpected signals. The available compute resources also put a limit to the bin selection and the number of pixels analyzed in each bin, because adding more bins and pixels will proportionally increase the runtime.

For the solution of problem in Eq.~\ref{eq3}, we use a sequential approach by first recovering $p$ with an exhaustive search, and then recovering $\mathbold{s}$ with a non-negative least-squares solver \cite{lawson1995solving}. For the solution of problem in Eq.~\ref{eq4}, we use a cyclic coordinate-descent approach by iteratively searching along coordinates in a cyclic order: pitch, yaw, surge, heave, roll, sway. Alternatively, we can use a randomized order or an order based on Gauss-Southwell rule \cite{nutini2015coordinate} which may provide a faster convergence rate for certain problems. We use binary search algorithm for optimizing along each coordinate with an error tolerance of \SI{0.01}{\degree} for rotations and \SI{1}{\micro\meter} for positions. We stop the iterations when a maximum iteration number of 100 is reached or when the desired error tolerance is met. The corresponding convergence plots of the algorithm and the cost function reduction at each iteration are presented in Fig.~\ref{fig8}. Surge iterations converge after the first cycle of coordinates. Other parameters converged gradually after five cycles of coordinates. The cost reduction converges at a logarithmic rate. Increasing the binary search iterations may lead to a marginal improvement in the cost function, however its contribution to accuracy is often beyond our visual perception when the relative error reaches a threshold of $10^{-6}$. Also the numerical errors due to discretization and numerical precision can become significant beyond a certain level of accuracy tolerance of the algorithm.

\begin{figure}
\centering
\includegraphics{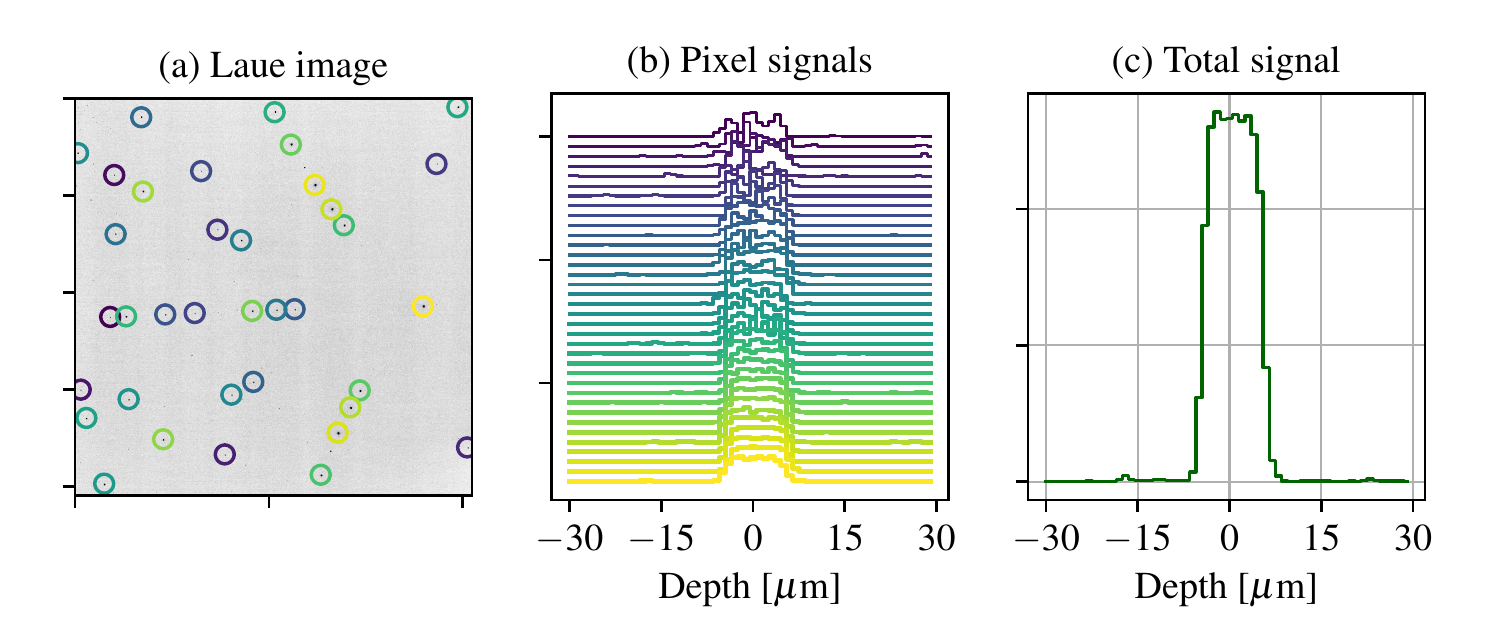}
\caption{Depth-resolved signals in focus. (a) One of the Laue images from the calibration dataset. Selected detector pixels for optimization are marked with color-coded circles. (b) Individual depth-resolved signals corresponding to the selected pixels. (c) Total signal obtained by summing all the depth-resolved signals. }
\label{fig9}
\end{figure}

Fig~\ref{fig9} shows the depth-resolved signals in focus along the illumination path. The pixels that we use for validation are marked with circles in Fig~\ref{fig8}-(a). The mean and standard variation of the recovered positions are respectively \SI{0.472}{\micro\meter} and \SI{0.552}{\micro\meter}. We observe that the position of the signals are estimated at sub-pixel accuracy ($\leq\SI{1}{\micro\meter}$) according to the median of their cumulative sums. We also see that some of the recovered signals do not resemble a boxcar function as we expect but they have fluctuations on top. This fluctuation can be considered as the signal reconstruction noise and it is directly related to the variable SNR in measurements, which lead to an inaccurate recovery of low SNR signals compared to the high SNR signals. However, the contribution of noisier signal reconstructions to the sum of signals is proportional to the diffracted intensity, so they induce a marginal error in the overall signal reconstruction. The total signal is obtained by summing all the depth-resolved signals from individual pixel recordings is plotted in Fig~\ref{fig8}-(c). It has a reasonably flat-top and its full-width at half-maximum is measured to be \SI{10}{\micro\meter} and it is matching with the thickness of the calibration sample.

\subsection{Evaluation with experimental data}
 
\begin{figure}[t]
\centering
\includegraphics{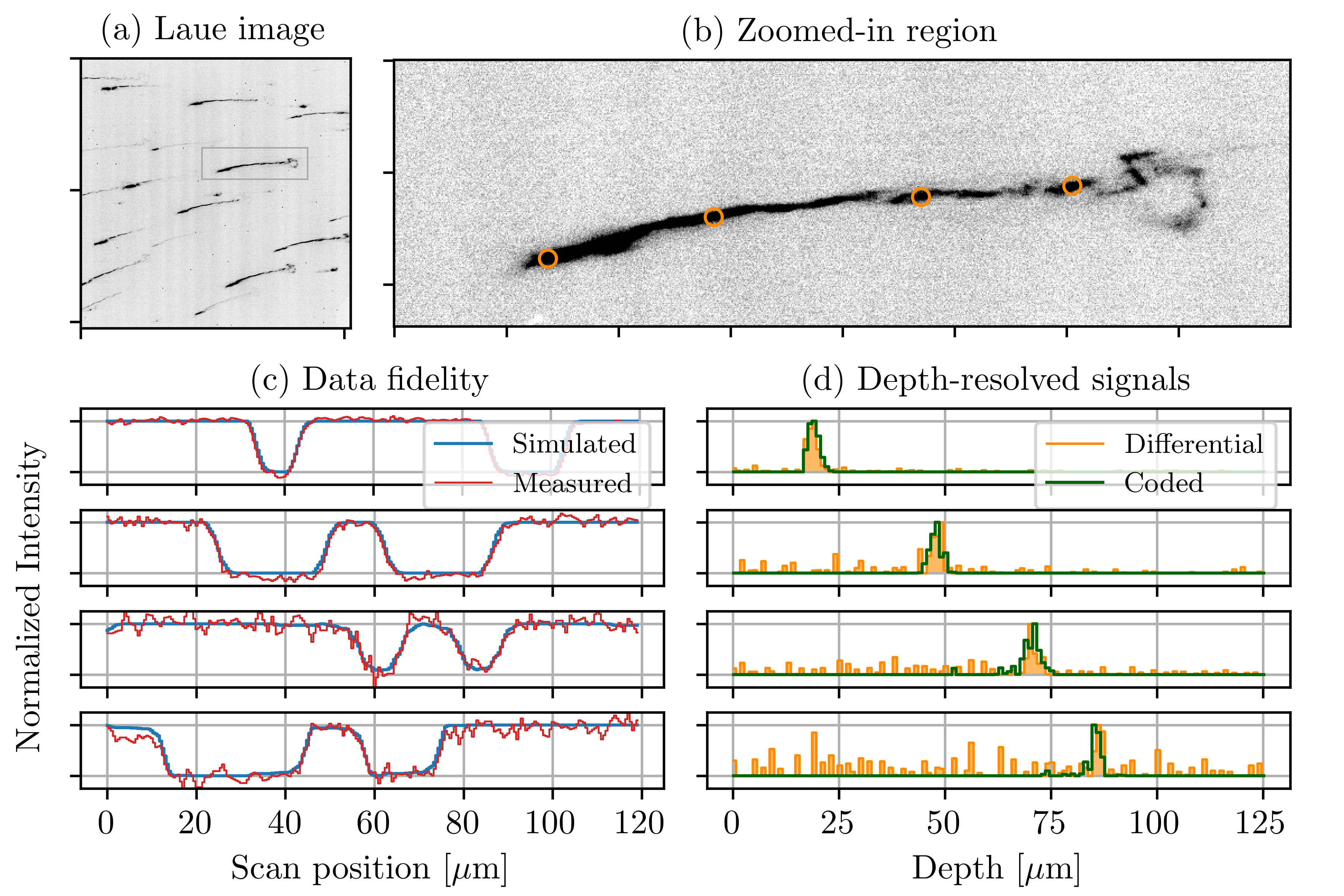}
\caption{Validation of estimated geometry with experimental data. (a) Measured Laue diffraction image that contains all the Bragg spots along the x-ray beam. (b) The zoomed-in image show the locations of the four pixels that are used for the analysis. (c) Measured and simulated intensities in the marked pixels in the zoomed-in image. (d) Depth-resolved signals for the marked pixels using the differential-aperture and the coded-aperture imaging methods at the same \ce{Ni} sample location.}
\label{fig10}
\end{figure}

We evaluate the reproducability of the geometry with scan data acquired from another sample. We use a heavily deformed polycrystalline nickel (\ce{Ni}) foil with a thickness of about \SI{75}{\micro\meter} as the imaging target. The sample was obtained by cutting the \ce{Ni} foil with a scissor and we use an arbitrary region close to the cut for collecting diffraction data. We mount the sample at a \SI{45}{\degree} angle towards the incident x-ray beam, which gives sample depth range along the beam direction of about \SI{100}{\micro\meter}. We place the coded-aperture about \SI{1}{\milli\meter} above the sample and scan it again with a \SI{1}{\micro\meter} step size along the beam direction over a \SI{120}{\micro\meter} interval. We use the same scanning geometry that we estimate through digital autofocusing. We show the reconstruction results of recovering $p$ and $\mathbold{s}$ in Fig.~\ref{fig10} for four selected detector pixels at marked positions. We plot the simulated and measured signals based on the recovered $p$ and $\mathbold{s}$. We obtain the simulated signals by computing the matrix product of the coding matrix and $\mathbold{s}$, and align the measurement signals relative to their recovered positions $p$ on the coded-aperture. As a final step, we register signals along the beam path using ray-tracing analysis as similar to the differential-aperture imaging method \cite{Larson:02}. We validate our reconstructions with the conventional differential-aperture method (a \SI{100}{\micro\meter} platinum wire scanned with a \SI{1}{\micro\meter} step size across the sample) for the same \ce{Ni} sample location, and observe that the focusing parameters are yielding comparable depth-resolved signals with the differential-aperture method.
 
\begin{figure}[t]
\centering
\includegraphics{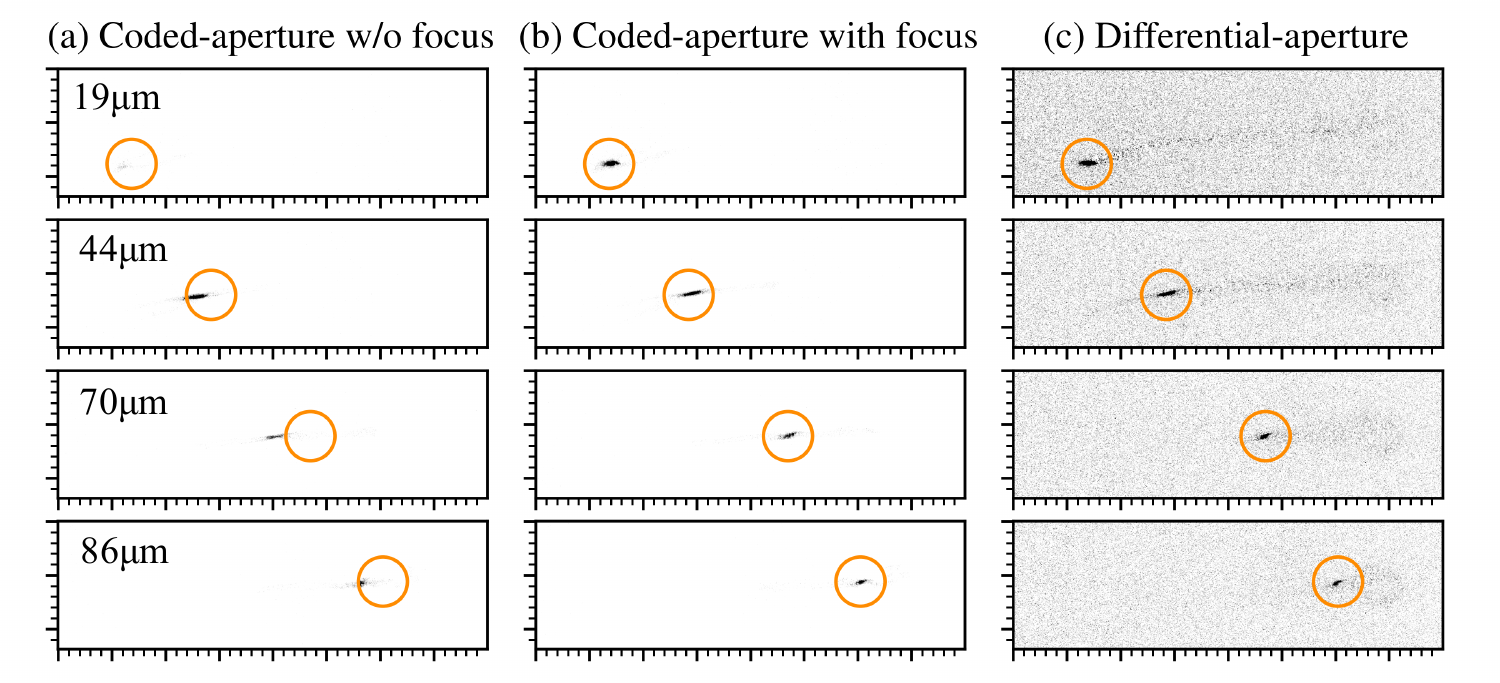}
\caption{Resolved signals for the marked pixels in Fig.~\ref{fig9}-(b) at depths 19, 44, 70 and \SI{86}{\micro\meter} using the coded-aperture with and without focus and the differential-aperture. Measurements for coded- and differential-aperture methods are recorded at the same \ce{Ni} sample location.}
\label{fig11}
\end{figure}

To validate the accuracy of the digital autofocusing, we present in Fig.~\ref{fig11} the depth-resolved signals obtained with and without the focused coded-aperture, together with the depth-resolved signals obtained from a differential-aperture scan as the baseline approach. We mark with circles the Bragg signals originating from a \SI{1}{\micro\meter} length at depths 19, 44, 70 and \SI{86}{\micro\meter} corresponding to the marked pixels in Fig.~\ref{fig9}-(b). We see that when the coded-aperture is out of focus, the Bragg spots shape and position can deviate from their true locations as probed by the differential-aperture. We observed a shift of diffraction peaks in the range of 20-40 pixels, which corresponds to 4-\SI{8}{\milli\meter} distance error in the detector plane and 8-\SI{16}{\milli\radian} angular deviation from its assumed direction from the source. However, the depth-resolved signals from an in-focus coded-aperture match nicely with the ones obtained from a differential-aperture. Note that the signals are recovered using 120 scan points in coded-aperture as compared to the 600 scan points of a differential-aperture based microscope, yielding 5 times speed-up in acquisition. In addition, the images are significantly less noisier with a coded-aperture, probably because of the multiple crossings of absorbing structures as compared to a single absorbing structure in a differential-aperture.

\section{Discussions}

Note that the columns of the coding matrix $\mathbold{A}_p$ in Eq.~\ref{eq1} presents an ideal raster-scan, but in reality, the step sizes of the linear translation of the coded-aperture can deviate slightly from the intended step size. For example, we experience deviations on the order of tens of nanometer during scanning. This effect is negligible for imaging samples at a micrometer resolution, but it may affect the reconstruction quality when a resolution finer than \SI{100}{\nano\meter} is desired. A straightforward approach to lift this limitation is to use an interferometer system to measure the position of the sample with respect to the linear motors that are used to translate the coded-aperture. While this approach is required for nanoscale imaging, it is also computationally more demanding because of the added cost of discretization of the coded-aperture function that involves sub-pixel interpolations to calculate the coefficients of the coding matrix. 

Prior to x-ray experiments, we can use electron and optical microscopy techniques to characterize the structure of the coded-aperture. While these techniques can provide a good precision in a local area, characterizing a large area at high resolution requires stitching of many images at a sub-pixel accuracy, which is challenging. The main challenge is to keep the coded-aperture within the depth-of-field of the microscope such that we don't have blurry images which hinders the accuracy of sub-pixel registration. In addition, they only provide information about the surface of the coded-aperture and lacks quantification of the thickness across the coded-aperture plane. An alternative approach is to use our digital autofocusing method for a self-characterization of the coded-aperture with the same x-ray microscope that we use to acquire data. We can model the physical geometry of the coded-aperture with parameters such as thickness or bar sizes with their intended locations, and optimize parameters until the simulated coded-aperture structure is matching with the real one generating the calibration data. The advantage of this approach is that we characterize the overall structure only from the areas where the diffracted beams pass through. 

We use the variance of individual depth-resolved signal positions from the calibrated laboratory reference as the basis for constructing the cost function in Eq.~\ref{eq4}. However, other cost functions may as well provide a similar or better metric in certain conditions. For example, noisy measurements from a less absorbing coded-aperture can be compensated by adding extra terms to the data fidelity term for regularizing the depth-resolved signals. The same regularization approach can also be used to shorten the time to acquire the calibration data when the x-ray source is not stable over the time. When the source is unstable, we can collect a calibration dataset with a low exposure setting or with less number of scan points and use regularization techniques to address the measurement noise or less number of data points. Another approach is to weigh the contribution of each depth-resolved signal in the data fidelity term for added robustness. The weights can be calculated based on the measurement noise for resolving each signal. Last but not least, we can choose different measures for evaluating the data fidelity term. For example, choosing the energy of the depth-resolved signals (do not confuse it with the energy of the source or the diffracted beams) and maximizing their sum can provide another alternative. 

Autofocusing is a computationally intensive process. Optimization along a coordinate direction requires reassembling of the coding matrix and solving the two problems given in Eq.~\ref{eq3} and Eq.~\ref{eq4} many times, and we repeat this step many times by changing the coordinate direction till we reach a solution. For the results in this paper, we used recording from 36 pixels each corresponding to a different Bragg spot, and the processing took several hours on a 4.2 GHz Quad-Core Intel Core i7. Depending on the numerical algorithm and the processing hardware used for solving the signal reconstruction problem in Eq.~\ref{eq3}, the time to solve a pixel can range from a fraction of a second to a minute. For example, some algorithms such as based on factorization (e.g., LR, QR  or Cholesky) can be accelerated with graphical processing units \cite{agullo2009numerical}, and some linear solver like simplex methods are more challenging to scale. However, when an accurate but slow solver is used, there are paths we can follow to speed up the process. A straightforward approach is to leverage parallel computing by distributing the reconstruction problems to multiple processing cores. Because the data footprint of these two problems are small, the parallelism can provide a linear acceleration up to the number of pixels used in the analysis. In addition to brute parallelism, we can leverage the graphical processing unit implementations to accelerate suitable linear solvers for such problems, and the Euler-Rodrigues formula \cite{euler1776nova, rodrigues1840lois} for efficiently calculating the rotations of the coded-aperture. 

Our current approach can resolve individual signals diffracted into a detector pixel, but in rare cases, two beams can diffract to the same detector pixel. Resolving two or more signals from a single pixel recording is more challenging and requires additional modeling. While this types of interference events deserves additional modeling, for calibration purposes we can select the sample and design the experiments such that no interference happens.

Redundancy in calibration data is useful to combat with measurement noise and other unknown experimental factors. Therefore, a good practice is to scan the full extent of the coded-aperture instead of the bare minimum needed to resolve a depth (e.g., 8-bit sequence for our aperture). With our experimental setup, this corresponds to collecting about 2500 images and takes about less than an hour with the current APS source. A calibration data from the full extent of the aperture is also useful if we want to evaluate all the bar widths of the coded-aperture, uniformity in thickness, and bending if there are any.

The scanning geometry that we used in the paper for autofocusing is probably the best option for autofocusing because it provides the sharpest intensity modulation in the detector. Because we have full control on the calibration sample, we can avoid the risk of collision by imaging a region close to the edge of the calibration sample. However, for imaging samples with arbitrary shapes, scanning the coded-aperture at a \SI{45}{\degree} tilted geometry along the surface of the sample can yield higher resolution, because the distance from the coded-aperture to the sample surface sets the limit to achievable resolution. While in a tilted geometry the modulation will be tapered, the algorithm can still provide reconstructions given that the coding matrix is constructed according to the tilted geometry.

Beam energy will affect the intensity modulation contrast in measurements. When all the experimental parameters are kept the same, a higher beam energy will lead to a reduced modulation contrast due to lower absorption of the diffracted beams as they traverse the coded-aperture. However, we can correspondingly increase the exposure time to surpass the measurement noise relative to the peak-to-peak amplitude of the modulation. When the detector saturation due to long exposure sets a limit to allowable exposure for taking an image, we can collect multiple images and use the sum the images to improve the statistics. As an alternative or a complementary approach, we can use thicker coded-apertures that can absorb more photons. However, while a lower energy will lead to a better contrast, our linear approximation to beam attenuation can also induce artifacts (especially for non-zero incidence angles) in resolved signals. One can use non-linear models to more accurately capture the exponential dependency between the beam attenuation and the traversed thickness of the beam.

We use a smoothed boxcar function to approximate an initial state of the signal $\mathbold{s}$ and it provided satisfactory estimates for the experiments that we have performed so far. This is probably because the crystallites are locally textured and not arbitrarily distorted. For the cases, when more complex signals are involved, we may employ bootstrapping approaches that can be used to obtain confidence intervals for the sample shape at the cost of additional computation. This requires solving the reconstruction problem many times with randomly generated signals from a distribution and use the signals that yield the minimum discrepancy between the simulated and raw data.
 
As a conclusion, we introduce the geometrical model of a coded-aperture Laue diffraction microscope, and propose a method to focus the microscope digitally using a calibration data from a known sample. This study provides a deeper insight about the effect of coded-aperture positioning and scanning geometry on the reconstruction performance, and suggests alternative ways to improve the focusing with different algorithms and data acquisition schemes.

\section*{Acknowledgements} This research used resources of the Advanced Photon Source and the Center for Nanoscale Materials, U.S. Department of Energy (DOE) Office of Science User Facilities and is based on work supported by Laboratory Directed Research and Development (LDRD) funding from Argonne National Laboratory, provided by the Director, Office of Science, of the U.S. DOE under Contract No. DE-AC02-06CH11357.

\bibliographystyle{ieeetr}
\bibliography{main}

\end{document}